\documentclass[acus]{JAC2000}

\usepackage{graphicx}

\begin{document}
\title{THE SPALLATION NEUTRON SOURCE:\\ 
A POWERFUL TOOL FOR MATERIALS RESEARCH
\thanks{Work supported by the U.S. Dept. of Energy 
under contract DE-AC05-00OR22725}}
\author{T.E. Mason, T.A. Gabriel, R.K. Crawford, K.W. Herwig, 
F. Klose, and J.F. Ankner,\\ SNS Project, 701 Scarboro Rd. Oak Ridge,
TN 37830}
\maketitle

\begin{abstract} 
The wavelengths and energies of thermal and cold neutrons are ideally
matched to the length and energy scales in the materials that underpin 
technologies of the present and future: ranging from semiconductors to 
magnetic devices, composites to biomaterials and polymers. The
Spallation Neutron Source (SNS) will use an accelerator to produce the most
intense beams of neutrons in the world when it is complete at the end
of 2005. The project is being built by a collaboration of six
U.S. Department of Energy laboratories. It will serve a diverse
community of users drawn from academia, industry, and government labs
with interests in condensed matter physics, chemistry, engineering
materials, biology, and beyond.
\end{abstract}

\section{INTRODUCTION}
With materials of ever increasing complexity becoming key elements of
the technologies underpinning industrial and economic development
there is an ongoing need for tools that reveal the microscopic origins
of physical, electrical, magnetic, chemical, and biological
properties.  Neutron scattering is one such tool for the study
of the structure and dynamics of materials\cite{MRS}. Neutrons are well suited
to this purpose for several reasons:

\begin{itemize}
\item Neutrons are electrically neutral leading to penetration depths 
of centimeters;
\item Neutron cross sections exhibit no regular dependence on atomic 
number and are similar in magnitude across the periodic table giving
rise to sensitivity to light elements in the presence of heavier ones;  
\item The range of momentum transfer available allows probing of a 
broad range of length scales (0.1 to $10^5 \AA$) important in many
different materials and applications; 
\item Thermal and cold neutrons cover a range of 
energies sufficient to probe a wide range of lattice or magnetic 
excitations;
\item Neutrons have magnetic moments and are thus uniquely sensitive 
probes of magnetic interactions; 
\item Neutrons can be polarized, allowing the cross-sections(magnetic 
and nonmagnetic) to be separated;
\item The simplicity of the magnetic and nuclear interactions make 
interpretation of results straightforward.
\end{itemize}

Advances in neutron scattering have, from its earliest days, been
driven by the scientific opportunities presented by improved source 
performance and instrumentation optimized to take advantage of that 
performance.  The Spallation Neutron Source represents a substantial
advance in neutron source performance over any facility in the world
and, together with improved instrumentation, will make possible
measurements of structure and dynamics with unprecented intensity,
resolution, and dynamic range.  The accelerator complex that is
the proton driver for the SNS is described elsewhere in these 
proceedings.  What follows is a brief summary of the Experimental
Facilities that will turn the protons into science.

\section{SNS EXPERIMENTAL FACILITIES}

\subsection{Target Systems}
\begin{figure}[hbt]
\centering
\includegraphics*[width=80mm]{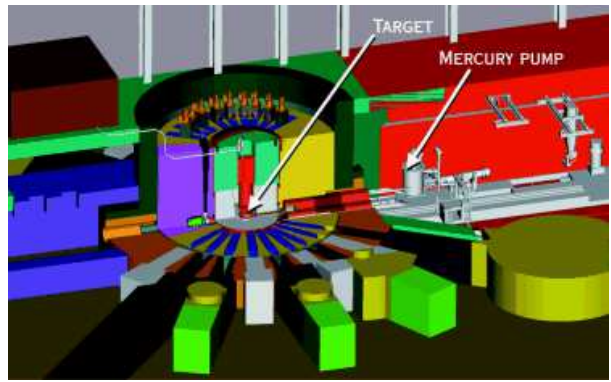}
\caption{Cross-sectional view of the bulk shielding, shutters,
instruments, maintenance cells, and mercury target.}
\label{targetstation}
\end{figure}

The development of the SNS Target System, including preliminary design
and research and development, is progressing as planned. A
global overview of the target/instrument hall is shown in 
Fig.~\ref{targetstation}. The 
mercury-based target system, which will receive the short-pulsed,
proton beam (2 MW, 1-GeV protons, 60 Hz, $<1.0$ $\mu$s/pulse), will be
located in the center of the bulk shielding. As shown in 
Fig.~\ref{targetstation}, the 
shielding is slightly left of center, with the maintenance cells to
the right. The first maintenance cell will monitor and maintain the
mercury process and pumping system. Additionally, because the system
is designed to require only five days for replacement of the stainless
steel (SS316L) target container and because the first maintenance
cell will allow replacement every six weeks, a good availability of
neutrons will be maintained. The remaining cells will be used for
storage, maintenance, and shipping of various components such as
shutters, the proton beam window, and the inner core plug that holds
the moderators.

\begin{figure}[htb]
\centering
\includegraphics*[width=80mm]{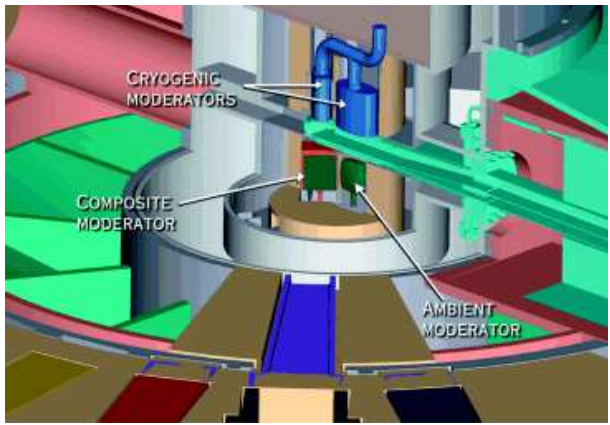}
\caption{Cross-sectional view of the cryogenic, composite, and ambient
moderators and the mercury target.  The proton beam arrives from the
upper left of the drawing.}
\label{target_closeup}
\end{figure}

A sliced section of the target, moderator, and reflector system is
shown in Fig.~\ref{target_closeup} . Cryogenic moderators 
[supercritical light hydrogen 
(H2) @ 19 K] are located on the top of the target. The upstream bottom 
moderator will contain a composite of supercritical light hydrogen and
light water. The purpose of this composite is to increase the number
of neutrons in the 100 K range. Neutrons in this energy range can be
used for a multitude of applications. Preliminary analysis indicates
that this increase could approach the neutron yield expected from a
liquid methane moderator.  The downstream bottom moderator is ambient
light water.

\begin{figure}[hbt]
\centering
\includegraphics*[width=80mm]{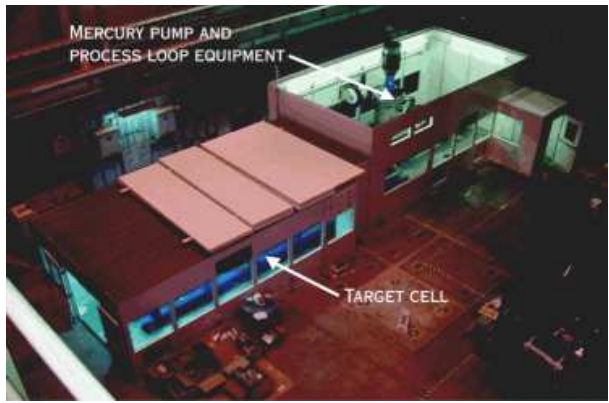}
\caption{Target test facility showing the actual location of the
target cell, mercury pump, and process loop equipment.  In the actual
facility the proton beam would be coming from the left
side.}
\label{ttf}
\end{figure}

In support of the target design, we have focused our R\&D program on
the development of a mercury target system and have included research
in thermal shock, thermal hydraulics, material damage and
compatibility, and remote handling. All of our large R\&D mercury loops
are now operational. The use of these loops will yield valuable
information on thermal hydraulics, remote handling, and operator
requirements and training. The enclosure structure and part of the
mercury loop structure are shown in Fig.~\ref{ttf}. This structure is a
prototypical mercury loop that will be similar to the one used in the
actual SNS facility. The target end is located in the left side of the 
enclosure. The motor and sump pump can be seen on the right side. The
system, which holds about 20 metric tons of mercury, became
operational on October 31, 1999, and reached its design requirements
shortly after startup. At the
maximum pump speed of 600 rpm, the mercury flows at a rate of ~30 l/s.

\subsection{Instrument Systems}
\begin{figure}[hbt]
\centering
\includegraphics*[width=80mm]{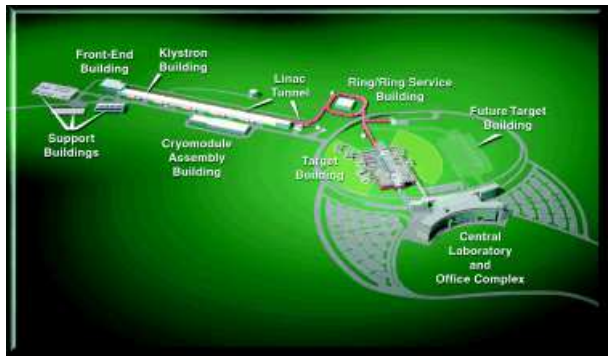}
\caption{Spallation Neutron Source site plan showing the front end,
linac, accumulator ring, target, instruments, and central lab/office
and other support facilities.}
\label{site_plan}
\end{figure}

When SNS is complete and operating at 2 MW, it will offer
unprecedented performance for neutron-scattering research, with more 
than an order of magnitude higher flux than any existing facility. 
Fig.~\ref{site_plan} shows the layout of the SNS accelerator complex,
experimental facilities, and support facilities.
To realize the potential this offers for research in chemistry,
condensed matter physics, materials science and engineering, and 
biology, a world-class suite of instruments is being developed that 
makes optimal use of the SNS beams and that is suited to the needs of 
users across a broad range of disciplines. The mechanisms by which 
instruments are built and operated will be responsive to varying 
degrees of experience, from new graduate students and first-time 
neutron users to experienced users with an interest in instrument 
design. As a DOE user facility, the SNS construction budget 
includes funds for an initial suite of instruments that will be 
available to users through a peer-reviewed proposal system. These 
instruments are being selected in consultation with the user 
community, following advice from our 
Instrument Oversight Committee (IOC), 
and are being designed in consultation with prospective users 
through Instrument Advisory Teams (IATs). We expect that 75\% 
of the beam time on this instrument suite will be available 
through the proposal system, with the remaining 25\% for use 
by in-house scientific staff, for testing and calibration, for 
feasibility studies by users before submission of proposals,  and 
for rapid response experiments that occur outside the regular proposal 
schedule. 

In addition to providing a fast start on instruments for SNS users, 
the initial instrument suite will allow the development of core 
technologies such as choppers, data acquisition systems, and control 
software that will form the basis for similar systems in other
instruments.  These systems will be standardized where appropriate to 
simplify use and maintenance and to reduce costs. Full 
"target-to-detector" computer simulation is being developed to
optimize and integrate target and instrument design. In many cases, 
the SNS will require instrument technologies beyond the current state 
of the art. Consequently, our construction budget for instruments is 
supplemented by a significant R\&D program, which will allow
development of new technologies that will form the basis for the 
initial instrument suite and provide room for growth.
Although the current instrumentation budget allows for 10 or 11 
best-in-class neutron-scattering instruments, a total of 24 can be 
accommodated on the high-power target station. Over time, new
instruments will be built for the additional beam lines as part of the 
normal operating life of the SNS. However, to achieve full utilization
of SNS, with the possibility of serving the focused research needs of 
groups willing to commit to building and operating neutron
instruments, it is desirable to provide for instrumentation built by 
Instrument Development Teams (IDTs) that may or may not include SNS as 
a member. IDTs would provide at least partial funding for an
instrument and would receive dedicated beam time in return for their 
financial commitment. For an instrument fully funded (including
operation) by the IDT, up to 75\% of beam time could be reserved for
the IDT, with the remainder open to general users.
The basic principles by which instruments are approved for SNS are the
same, whichever mode of access is involved. The main criteria for
instrument selection are the scientific program and the need for the 
unique capabilities of the SNS. The SNS is committed to seamless user 
access and instrument optimization across the facility. Instruments 
should be built on the beam line that best suits their requirements, 
and access for users should be uniform across the facility,
independent of the funding source for the instrument. Guidelines for 
instrument team proposals that will facilitate fair and systematic
evaluation are being developed in consultation with our advisory 
committees as well as the broader user community. 

The instrument team proposal process will advance in two phases: an 
initial letter of intent followed by a detailed proposal. The letter 
of intent will broadly outline the proposal with sufficient detail for 
evaluating the scientific potential, funding mechanism(s), and 
management plan. An approved letter of intent would be followed by a 
more detailed proposal. The IOC, as well as expert review, will be 
involved at the appropriate stage. Similar guidelines would be
followed for other (nonscattering) uses of the SNS facilities, with 
the modification that review would be by the Scientific Advisory 
Committee, supplemented by subject matter experts (because the IOC 
is a neutron-scattering expert panel). Such potential uses of the 
SNS would also be subject to the condition that they not compromise 
the neutron-scattering mission of the SNS and that the funding must 
be incremental to the project and cover all incurred costs. In 
addition to project-funded instruments (with an IAT) and totally 
externally funded instruments (with an IDT), there is a possibility 
for hybrid arrangements that would be negotiated on a case-by-case 
basis with the understanding that funding level and dedicated beam
time are commensurate.  

\begin{figure}[hbt]
\centering
\includegraphics*[width=80mm]{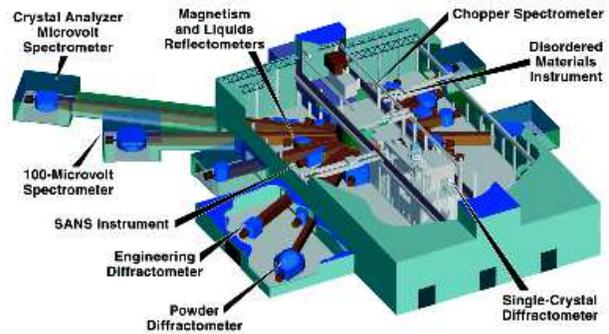}
\caption{Schematic instrument suite for the SNS. The actual
instrumentation will be determined in consultation with the
user community.}
\label{instrument_plan}
\end{figure}

There are currently three instruments that 
have been recommended for construction as part of the SNS Project.
These are shown in Fig.~\ref{instrument_plan} together with other
instruments under consideration and are described briefly below.

\subsubsection{High Resolution Backscattering Spectrometer}
\begin{figure}[hbt]
\centering
\includegraphics*[width=80mm]{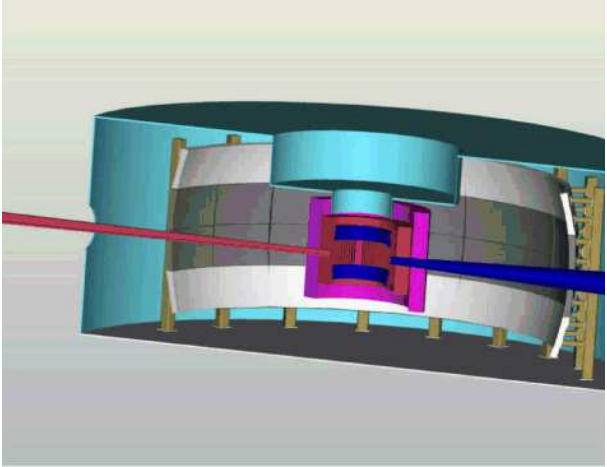}
\caption{View of the scattering chamber of the backscattering
spectrometer.  With the analyzer crystals located 2.5 m from the
sample, the chamber will have a diameter of approximately 6 m.}
\label{back}
\end{figure}

A mechanism for achieving high energy resolution in inelastic
scattering is to make use of Bragg scattering close to the
backscattering condition to determine the final wavelength of
neutrons scattered from the sample.  With a pulsed source,
time-of-flight is used to determine the incident neutron energy.
\cite{gehring97,frick97,carlile92}
The SNS instrument is a near-backscattering, crystal-analyzer
spectrometer designed to provide extremely high-energy resolution 
($\hbar\omega = 2.2 \mu$eV FWHM, elastic). The design requires a long 
initial guide section of 84 m from moderator to sample to achieve the 
timing resolution necessary to achieve the desired $\delta\omega$. 
The scattering chamber design is illustrated in Fig.~\ref{back}. 
Neutrons focused onto the sample by the supermirror funnel scatter 
towards the analyzer crystals. The strained, perfect silicon (111) 
crystals reflect neutrons with a narrow distribution of energies
centered at $2.082 \mu$eV onto the detectors. The design is optimized 
for quasi-elastic scattering but will provide 0.1\% resolution in 
energy transfer $\hbar\omega$, up to $\hbar\omega = 18$ meV. This spectrometer 
will provide an unprecedented dynamic range near the elastic peak of 
$-258 \mu eV < \hbar\omega < 258 \mu eV$, about seven times that of 
comparable reactor-based instruments. For experiments that require 
the full dynamic range available at reactor-based instruments 
(or greater), we expect this spectrometer to have an effective 
count rate of ~100 times that of the current best spectrometers.

\subsubsection{Reflectometers}
\begin{figure}[hbt]
\centering
\includegraphics*[width=80mm]{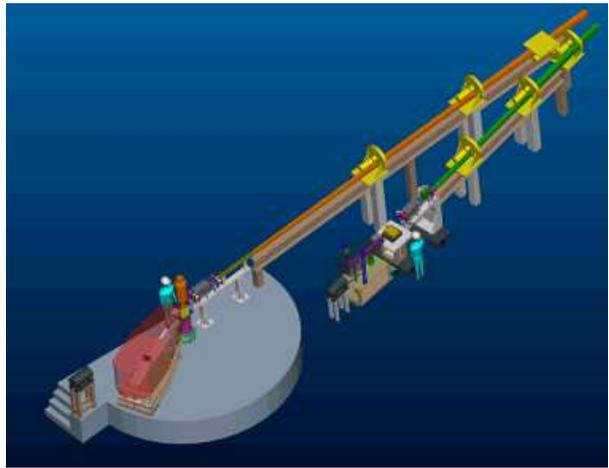}
\caption{Three-dimensional representation of the reflectometry
beamlines, viewed from the detectors looking towards the cryogenic
moderator.
On the left is the polarized beam reflectometer (vertical surface), on
the right the liquids reflectometer (horizontal surface)}
\label{refl}
\end{figure}

Neutron reflectometry is the study of elastic scattering from 
surfaces and interfaces at glancing angles.  It provides structural
information about surface and near surface atomic and magnetic order.
\cite{ankner99}
Two reflectometers are being considered for installation on a single 
beam line at SNS, one featuring an incident polarized beam for the
study of magnetic materials and one with a horizontal sample surface 
to facilitate the study of liquids (see Fig.~\ref{refl}). 
Both instruments will use advanced 
neutron optics. Supermirror-coated microguide beam benders eliminate 
fast-neutron and gamma backgrounds. Tapered supermirror guides
transport high flux to the sample position. The incident optics and 
bandwidth chopper system deliver ( $\lambda > 2.5 \AA$) neutrons to the 
sample at repetition rates of 60, 30, or 20 Hz. Running at 60 Hz, the 
instruments will be capable of measuring reflectivities of R $< 10^{-9}$, 
an order-of-magnitude improvement over the best existing instruments. 
Similar or greater improvements in data-collection rates have exciting 
implications for kinetic studies.

The polarized-beam reflectometer employs a vertical sample geometry to 
accommodate large superconducting magnets and other ancillary
equipment. In addition to collecting data in reflection geometry, the 
instrument will have a detector bank at high angles for diffraction
studies. Use of a Drabkin-flipper-type beam conditioning device and 
different polarizer, analyzer, and spin-flipper options are the
objects of a vigorous R\&D effort.

The liquids instrument features a novel design that uses the broad 
angular dispersion produced by the tapered guide. By sampling
different incident angles (5-15) with beam-defining slits and using 
the relatively narrow wavelength bandwidth available at 60 Hz, we 
can efficiently cover a large range of momentum (hQ) transfer. 
Operation in this mode uses all of the source flux and combines the 
counting efficiency of a fixed-wavelength reflectometer with the wide 
Q coverage of a broadband instrument.

\section{CONCLUSIONS}

The Spallation Neutron Source will be the world's leading facility for 
studies of the structure and dynamics of materials using thermal and
cold neutrons.  It leverages state of the art accelerator technology,
to deliver a high power (more than a factor of 12 times ISIS,
currently the world's most intense pulsed spallation source), high
reliability tool to physicists, chemists, biologists, and engineers.
When it is fully instrumented it will support 1000-2000 users per year
drawn from universities, industry, and government laboratories in the
U.S. and abroad.  Ultimately it will support two target stations,
operating
at different frequencies and power levels, doubling the capacity and
more than doubling its scientific performance.

\end{document}